# Disruptive Behavior Disorder (DBD) Rating Scale for Georgian Population

Vera Bzhalava[1], Ketevan Inasaridze[2]

[1]First Called Georgian University of the Patriarchate of Georgia

[2]Georgian State University of Physical Education and Sport

**Abstract**

In the presented study Parent/Teacher Disruptive Behavior Disorder (DBD) rating scale based on the Diagnostic and Statistical Manual of Mental Disorders (DSM-IV-TR [APA, 2000]) which was developed by Pelham and his colleagues (Pelham et al., 1992) was translated and adopted for assessment of childhood behavioral abnormalities, especially ADHD, ODD and CD in Georgian children and adolescents. The DBD rating scale was translated into Georgian language using back translation technique by English language philologists and checked and corrected by qualified psychologists and psychiatrist of Georgia. Children and adolescents in the age range of 6 to 16 years (N 290; Mean Age 10.50, SD=2.88) including 153 males (Mean Age 10.42, SD= 2.62) and 141 females (Mean Age 10.60, SD=3.14) were recruited from different public schools of Tbilisi and the Neurology Department of the Pediatric Clinic of the Tbilisi State Medical University. Participants objectively were assessed via interviewing parents/teachers and qualified psychologists in three different settings including school, home and clinic. In terms of DBD total scores revealed statistically significant differences between healthy controls (M=27.71, SD=17.26) and children and adolescents with ADHD (M=61.51, SD= 22.79). Statistically significant differences were found for inattentive subtype between control (M=8.68, SD=5.68) and ADHD (M=18.15, SD=6.57) groups, for hyperactive/impulsive subtype ADHD group (M=17.39, SD=6.74) and healthy controls (M=8.12, SD=6.07), for oppositional defiant disorder in healthy control (M=6.68, SD=4.59) and persons with ADHD (M=13.7, SD=5.82), for conduct disorder between healthy group (M=2.51, SD=2.99) and group with ADHD (M=8.31, SD=6.14). In general it was shown that children and adolescents with ADHD had high score on DBD in comparison to typically developed persons. In the study also was determined gender wise prevalence in children and adolescents with ADHD, ODD and CD. The research revealed prevalence of males in comparison with females in all investigated categories.

This research was supported by a grant N 021-08 from the Rustaveli National Science Foundation.

Correspondence concerning this article should be addressed to Ketevan Inasaridze

E-mail: kate_inasaridze@hotmail.com

The Disruptive Behavior Disorders (DBDs) especially oppositional defiant disorder (ODD) and conduct disorder (CD) are frequently co-occurring psychiatric disorders in approximately half of children and adolescents with attention deficit hyperactivity disorder (ADHD) (Waschbusch 2002; Connor et al. 2010, Anderson and Kiehl, 2013; Masi., 2015). The results of many researches showed that the development of DBDs are associated with negative interaction of psychosocial and neurobiological factors (Lavigne et al., 2016; Kerekes et al., 2014; Boden et al., 2010).

ADHD is one of the most common neurodevelopmental disorder which is characterized by developmentally inappropriate levels of innattention, locomotor hyperactivity and impulsivity (Faraone et al., 2015). According to the DSM-IV-TR three subtypes of ADHD are defined: ADHD combined subtype (ADHD-C), ADHD predominantly hyperactive/impulsive subtype (ADHD-PH) and ADHD predominantly inattentive subtype (ADHD-PI). The validity of differentiating of these subtypes is rather debatable.

Disruptive Behavior Disorders (DBDs) might best be described along a continuum as the emergence of ODD may be a precursor to CD (Turgay, 2009; Rowe et al., 2010). According to DSM-IV children with ODD are characterized by angry/irritable mood, argumentative/defiant behavior, or vindictiveness which lasts at least six months (APA, 2000). Persons with oppositional defiant disorder do not reveal aggression toward people and animal, do not destroy property, and do not show a pattern of theft or deceit (Campbell et al., 2000; Nolen-Hoeksema, 2014).

Conduct disorder (CD) is a behavioral and emotional disorder diagnosed in childhood/adolescence that presents itself through a repetitive and persistent pattern of behavior in which the basic rights of others or major age-appropriate norms are violated (Lahey et al., 2003; Baker, 2013). These behaviors are often referred to as "antisocial behaviors" (Hinshaw SP, Lee SS., 2003). It is often seen as the precursor to antisocial personality disorder, which is not diagnosed until the individual is 18 years old (APA, 2000).

The frequency of ODD among boys is higher prior to puberty, although the tendency does not persist after puberty period (Demmer et al., 2017). The CD is also more common in boys in comparison to girls (Nordström et al., 2013). The manifestation of CD is also different between males and females. CD onset in girls is generally prior to adolescence (Olino et al., 2010).

Long term outcomes of children and adolescents for establishing or excluding comorbidity of ADHD, CD, ODD and also distinguishing of subtypes of ADHD strong differential diagnosis needs to be done. One of the common used screening scale in children and adolescents for abovementioned purposes is parent/teacher DBD rating scale which was developed by Pelham and his colleagues (Pelham et al., 1992) according to DSM –IV-TR diagnostic criteria (Molina et al., 1998; Pelham et al., 2005; Evans et al., 2013).The Disruptive Behavior Disorder rating scale consists of 45 items with response categories ranging from *not at all* (0) *to very much* (3) and takes approximately 10 minutes to administer. This rating scale includes 9 items related to symptoms of predominantly inattentive subtype of ADHD (ADHD-PI), 9 items for Hyperactive/Impulsive subtype of ADHD (ADHD-PH), 15 items for Conduct disorder (CD), and 8 items for oppositional defiant disorder (ODD). The DBD rating scale also has subscale for measuring combined type of ADHD (ADHD-C) (items, 9,18, 23, 27, 29, 34, 37, 42, 44, 1, 7, 12, 19, 22, 25, 30, 33 and 35) in children and adolescents. For the combined subtype of ADHD 6 or more items have to be selected from ADHD-PI subtype and 6 or more items – from ADHD-PH category. Items 10, 14, and 21 were not included for assessment of children/ adolescents' behavioral disorders since they are from DSM-III-R and are not included in the scoring for a DSM-IV diagnosis.

Attention-Deficit/Hyperactivity Disorder, Oppositional Defiant Disorder, or Conduct Disorder can be determined in two ways. The first method involves counting symptoms for each disorder using the Disruptive Behavior Disorders (DBD) rating scale. The second method includes comparing the target child's factor scores (e.g., using a 2 SD cutoff) on the DBD Rating Scale to established norms.

For empirical assessment of ADHD, ODD, CD, and comorbidity of these disorders in Georgian children and adolescents, Disruptive Behavior Disorder rating scale (Pelham et al., 1992) has been translated into Georgian language and adapted for Georgian population.

**Methods**

**Objectives**

The study was designed for preparation of a reliable and valid scale in Georgian language for assessment of neurodevelopmental disorders such as an ADHD, ODD and CD and establishment of norms for Georgian population. Therefore the aims of the research were: (1) Translation of the Disruptive Behavior Disorder (DBD) rating scale into Georgian language and (2) determination of psychometric properties of Georgian DBD rating scale.

Based on the aims the study was separated into two steps. The first step was translation of the Disruptive Behavior Disorder (DBD) rating scale. The second step related to collecting data and establishment of Georgian Disruptive Behavior Disorder (DBD) rating scale norms for Georgian population.

**Step 1**. In this step translation of Disruptive Behavior Disorder (DBD) rating scale was done by English language philologists and assessment of content equivalence between Georgian and English versions of Disruptive Behavior Disorder (DBD) rating scale was done by qualified two psychologists and one psychiatrist. For determining the authenticity of Georgian translation the scale was translated back into English language by two English language philologists who had not been involved in initial translation of the rating scale.

**Step 2**. Second step included determination of psychometric parameters of Georgian Disruptive Behavior Disorder (DBD) rating scale.

The data were analyzed by different descriptive and inferential statistical methods using Statistical Package for Social Sciences 20.0 (SPSS 20.0).

**Participants.** Georgian version of Disruptive Behavior Disorder (DBD) rating scale was administered on a sample of 290 children and adolescents with age range of 6-16 years (mean age 10.50, SD=2.87). The sample included 153 males (mean age 10.42, SD=2.62) and 137 females (mean age 10.60, SD=3.14). Participants were recruited from different public schools of Tbilisi and the Neurology Department of the Pediatric Clinic of the Tbilisi State Medical University (Table 1).

Table 1. The distribution of gender, age and ADHD subtypes for controls and children/adolescents with ADHD.

| Control type | Gender | N | Mean Age | Standard Deviation |
|---|---|---|---|---|
| control | girl | 106 | 11.09 | 3.203 |
|  | boy | 97 | 10.90 | 2.604 |
|  | Total | 203 | 11.00 | 2.926 |
| Hyperactive/Impulsive | girl | 7 | 8.86 | 3.237 |
|  | boy | 15 | 9.20 | 2.111 |
|  | Total | 22 | 9.09 | 2.448 |
| Inattentive | girl | 11 | 9.18 | 2.401 |
|  | boy | 16 | 9.56 | 2.250 |
|  | Total | 27 | 9.41 | 2.275 |
| Combined | girl | 13 | 8.69 | 1.437 |
|  | boy | 25 | 9.84 | 2.824 |
|  | Total | 38 | 9.45 | 2.479 |
| Total | girl | 137 | 10.60 | 3.138 |
|  | boy | 153 | 10.42 | 2.622 |
|  | Total | 290 | 10.50 | 2.874 |

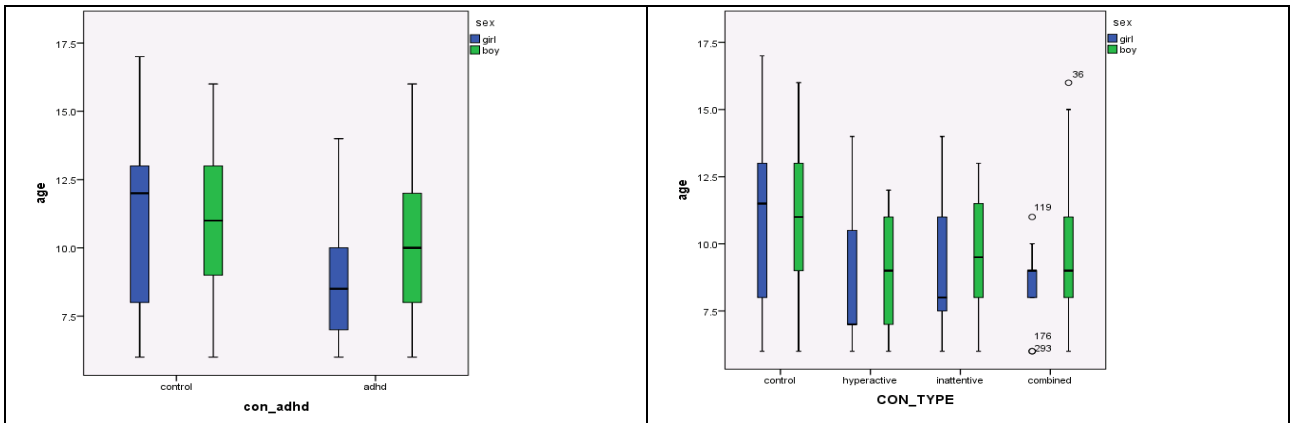

**Figure 1.** The distribution of gender, age and ADHD subtypes for controls and children/adolescents with ADHD

**Procedure**. Procedure. The study was conducted in agreement with Helsinki Declaration (1977) regarding ethical standards for clinical studies in medicine. Recruited participants were assessed by the qualified neuropsychologists/psychologists in three different settings including school, home and clinic. During assessing in school environment only those teachers were interviewed who were teaching children/adolescents participated in the research at least last one year. At home settings parents/caregivers of participated in the study children and adolescents were interviewed by psychologists. Raters were required to assess participants based on their behavior during the last six month in the appropriate settings. In clinic setting qualified neuropsychologists/psychologists based on their own observation objectively assessed participants.

**Results.**

In the presented study data were analyzed by different descriptive and inferential statistical methods using SPSS 20.0.

The mean scores and standart deviations were determined for typically developed and ADHD children and adolescents (Table 2). In terms of DBD total scores revealed statistically significant differences between healthy controls (M=27.71, SD=17.26) and children and adolescents with ADHD (M=61.51, SD= 22.79) (Fig. 2a). For the inattentive subtype significant difference was found for control (M=8.68, SD=5.68) and ADHD (M=18.15, SD=6.57) group (fig. 2b). For hyperactive/impulsive subtype statistically significant difference revealed between ADHD group (M=17.39, SD=6.74) and healthy controls (M=8.12, SD=6.07) (Fig. 2c). For oppositional defiant disorder in healthy control (M=6.68, SD=4.59) and persons with ADHD (M=13.7, SD=5.82) was found statistically significant difference (Fig. 2d). For conduct disorder revealed significant difference between healthy group (M=2.51, SD=2.99) and group with ADHD (M=8.31, SD=6.14) (Fig. 2e.). In general it was shown that children and adolescents with ADHD had high score on DBD in comparison to typically developed persons.

| variable | control | | ADHD | | variable | control | | ADHD | |
| --- | --- | --- | --- | --- | --- | --- | --- | --- | --- |
| | median | mode | median | mode | | median | mode | median | mode |
| DBD1 | 1.00 | 1 | 2.00 | 2 | DBD22 | .00 | 0 | 2.00 | 2 |
| DBD2 | .00 | 0 | .00 | 0 | DBD23 | 1.00 | 1 | 2.00 | 3 |
| DBD3 | 1.00 | 0 | 2.00 | 2 | DBD24 | 1.00 | 0 | 2.00 | 1 |
| DBD4 | .00 | 0 | 1.00 | 1 | DBD25 | .00 | 0 | 2.00 | 3 |
| DBD5 | .00 | 0 | .00 | 0 | DBD26 | 1.00 | 0 | 2.00 | 2 |
| DBD6 | .00 | 0 | .00 | 0 | DBD27 | 1.00 | 0 | 2.00 | 2 |
| DBD7 | 1.00 | 0 | 2.00 | 3 | DBD28 | 1.00 | 1 | 2.00 | 2 |
| DBD8 | .00 | 0 | .00 | 0 | DBD29 | 1.00 | 0 | 2.00 | 3 |
| DBD9 | 1.00 | 1 | 2.00 | 3 | DBD30 | 1.00 | 0 | 2.00 | 2 |
| DBD10 | .00 | 0 | 1.00 | 0 | DBD31 | .00 | 0 | .00 | 0 |
| DBD11 | .00 | 0 | .00 | 0 | DBD32 | .00 | 0 | 1.00 | 0 |
| DBD12 | 1.00 | 0 | 3.00 | 3 | DBD33 | 1.00 | 0 | 2.00 | 3 |
| DBD13 | .00 | 0 | 2.00 | 2 | DBD34 | .00 | 0 | 2.00 | 3 |
| DBD14 | .00 | 0 | 1.00 | 0 | DBD35 | .00 | 0 | 2.00 | 2 |
| DBD15 | .00 | 0 | 1.00 | 1 | DBD36 | .00 | 0 | .00 | 0 |
| DBD16 | .00 | 0 | .00 | 0 | DBD37 | 1.00 | 0 | 2.00 | 3 |
| DBD17 | 1.00 | 0 | 2.00 | 2 | DBD38 | .00 | 0 | 1.00 | 1 |
| DBD18 | 1.00 | 0 | 2.00 | 2 | DBD39 | .00 | 0 | 2.00 | 3 |
| DBD19 | 1.00 | 1 | 2.00 | 3 | DBD40 | .00 | 0 | .00 | 0 |
| DBD20 | .00 | 0 | 2.00 | 2 | DBD41 | .00 | 0 | .00 | 0 |
| DBD21 | 1.00 | 0 | 2.00 | 2 | DBD42 | .00 | 0 | 2.00 | 3 |

Table 2. The mean scores of Georgian DBD for Healthy controls and children and adolescents with ADHD

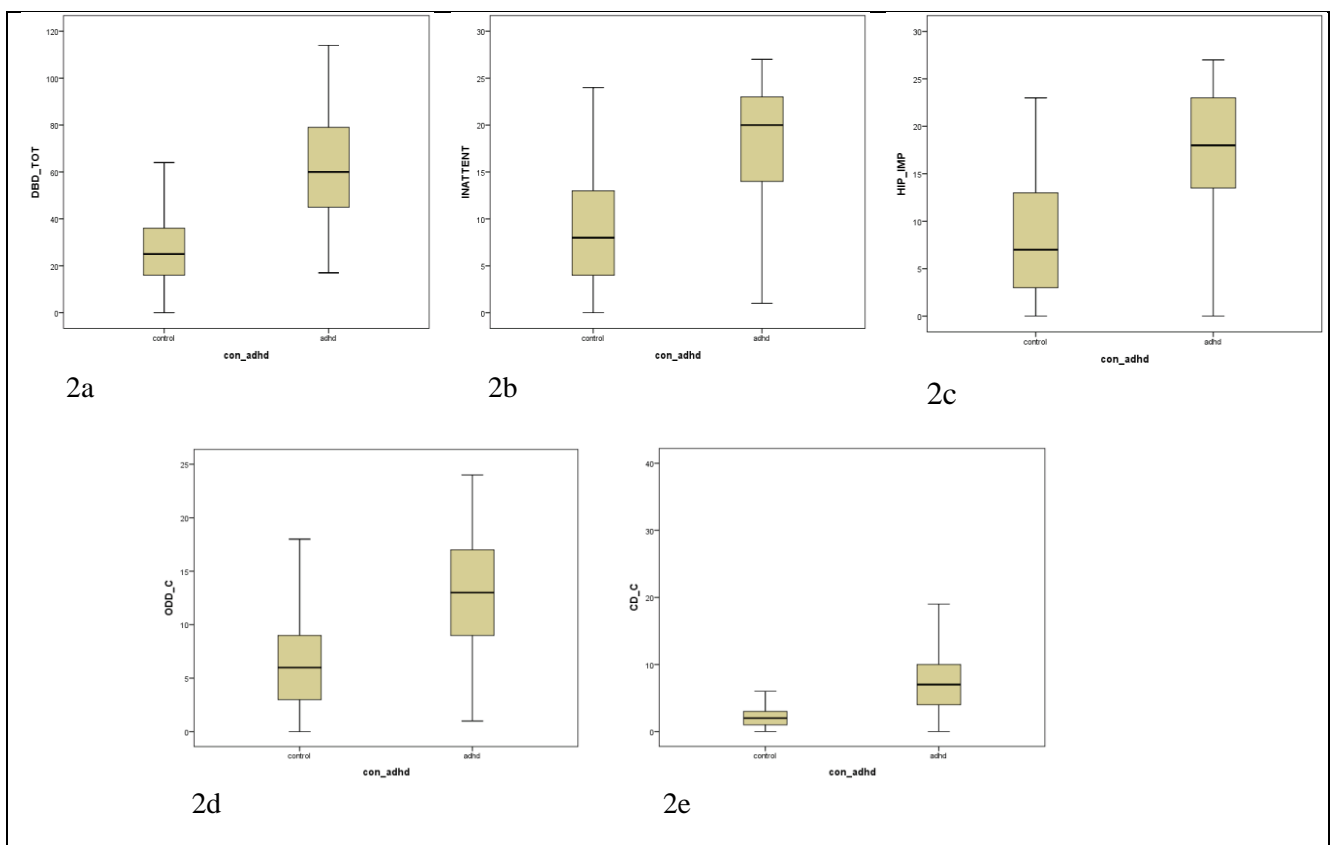

Figure 2. Measures and standard deviations of control and ADHD groups for Georgian DBD topics for 2a – total, 2b – inattentive, 2c – hyperactive/impulsive, 2d – oppositional defiant disorder and 2e – conduct disorder categories.

**Discussions**

The study was performed for preparing diagnostic scale for the assessment of children and adolescents with behavioral disorders such as ADHD, ODD and CD and also with comorbidity of these disorders in Georgia. For abovementioned purposes parents/teachers DBD rating scale was translated into Georgian language and psychometric properties of Disruptive Behavioral Disorder (DBD) rating scale were determined for Georgian population. As was shown in the study Georgian version of DBD rating scale can be successfully used for assessment of children's and adolescents' behavioral disorders both in educational and clinical settings. In general, it was shown that children and adolescents with ADHD had high score on DBD in comparison to typically developed persons. In the study also was determined gender wise prevalence in children and adolescents with ADHD, ODD and CD. The research revealed prevalence of males in comparison with females in all investigated categories. The Georgian version of DBD rating scale proved to be successfully applicable in the educational and clinical settings for screening and diagnosis for children and adolescents with behavioral disorders such as ADHD, ODD and CD.